\documentclass[12pt,document,nofootinbib,superscriptaddress, onecolumn,preprintnumbers,balancelastpage]{revtex4}
\pdfoutput=1
\usepackage{latexsym}
\usepackage{amssymb}
\usepackage{epsfig,amsmath,graphics}
\usepackage{color}
\usepackage{dcolumn}
\usepackage{slashed}
\usepackage{comment,latexsym} 
\usepackage{bm}
\usepackage{verbatim}
\usepackage{tabularx}
\usepackage{dcolumn}
\usepackage{hyperref}
\usepackage{setspace}
\usepackage{xspace}
\usepackage{epstopdf}

\def\be{\begin{equation}}
\def\ee{\end{equation}}
\newcommand{\beq}{\begin{equation}}
\newcommand{\eeq}{\end{equation}}
\def\bea{\begin{eqnarray}}
\def\eea{\end{eqnarray}}

\newcommand{\gsim}{ \mathop{}_{\textstyle \sim}^{\textstyle >} }
\newcommand{\lsim}{ \mathop{}_{\textstyle \sim}^{\textstyle <} }

\begin{document}

\begin{flushright}
\text{\normalsize MCTP-13-30}
\end{flushright}
\vskip 45 pt

\title{Neutralino Dark Matter with Light Staus}
\author{Aaron Pierce, Nausheen R. Shah, and Katherine Freese}
\affiliation{
Michigan Center for Theoretical Physics, \\
\vskip -6 pt 
Department of Physics, University of Michigan,\\
\vskip -6 pt
Ann Arbor, MI 48109}

\begin{abstract}
\vskip 15 pt
\begin{center}
{\bf Abstract}
\end{center}
In spite of rapid experimental progress, windows for light superparticles remain.  One possibility is a ${\mathcal O}$(100 GeV) tau slepton whose  $t$-channel exchange  can give the correct thermal relic abundance for a relatively light neutralino. We pedagogically review how this region arises and identify two distinct scenarios that will be tested soon on multiple fronts.  In the first, the neutralino has a significant down-type higgsino fraction and relatively large rates at direct detection experiments are expected.  In the second, there is large mixing between two relatively light staus, which could lead to a significant excess in the Higgs boson branching ratio to photons.  In addition, electroweak superpartners are sufficiently light that direct searches should be effective.
\end{abstract}


\maketitle
\newpage

\section{Introduction}
The neutralino of the minimal supersymmetric standard model (MSSM) represents an attractive dark matter candidate.  Collider searches have as yet yielded no sign of superpartners; for colored particles, some bounds now exceed a TeV.  Moreover, direct detection experiments are now rapidly pushing into the regime where signals of the lightest supersymmetic particle (LSP) of the MSSM might be observed.  Yet windows remain for light superpartners, particularly in models that relax theoretical assumptions such as gaugino mass unification.

In addition, the recent discovery of the Higgs boson \cite{Aad:2012tfa,Chatrchyan:2012ufa}, potentially opens a window onto the as yet hidden physics beyond the Standard Model. Intriguingly, initial measurements of Higgs boson decays indicated values close to Standard Model predictions for all channels except $h\to \gamma\gamma$, which was well in excess of the Standard Model expectation. Indeed, this channel is one where new physics might be expected to show itself, precisely because both the Standard Model and new potential physics  arise at the loop level. And while these measurements have moved towards the Standard Model value, there is still ample room for a non-trivial deviation.  At present $R_{\gamma \gamma}$, the ratio of the measured rate of Higgs decays to photons with respect to the Standard Model prediction, is measured by CMS to be 0.78-1.11~via two different analyses, with all observations consistent with the Standard Model at the $1\sigma$ level~\cite{CMS-PAS-HIG-13-001}. On the other hand, ATLAS still sees a significantly higher rate of 1.65, with a $\sim2.4\sigma$ deviation from the Standard Model~\cite{ATLAS-CONF-2013-012}.  If there were to be an excess in this channel, a natural interpretation is the existence of new charged particles who receive at least some of their mass from the Higgs mechanism, see, e.g. Ref.~\cite{Carena:2012xa}. A well motivated possibility is a light stau, the superpartner of the tau, a possibility studied recently in this context, for example, in Refs. \cite{Carena:2011aa, Carena:2012gp, Carena:2013iba, Cao:2012fz, Giudice:2012pf, Sato:2012bf, Haisch:2012re}. Indeed the stau is  one of the superpartners whose mass can be quite light ${\mathcal O}$(100 GeV), in spite of the progress of limits at the LHC.

If such a light stau were to exist, it could have implications for the thermal history of dark matter, as has been shown in recent scans performed over the MSSM parameter space~\cite{AlbornozVasquez:2011yq,CahillRowley:2012cb, Arbey:2012na, Haisch:2012re, Han:2013gba,Cahill-Rowley:2013dpa, Belanger:2013pna, Fowlie:2013oua, Boehm:2013qva, Choudhury:2012tc}, see also Ref.~\cite{Calibbi:2013poa}. Our interest here is not in the stau-coannihilation region \cite{Ellis:1999mm}, where a large enough annihilation cross section can be achieved if the dark matter and the stau are nearly degenerate.   Instead our focus is on the ``bulk stau'' region where no particular mass relationship is required -- rather a  light stau acts as a  $t$-channel mediator to ensure the proper relic density for a predominantly bino LSP (via the process $\chi \chi \rightarrow  \tau \tau$).  This is to be contrasted with the selectron and the smuon,  as reviewed in Ref.~\cite{ArkaniHamed:2006mb}, where LEP limits \cite{Heister:2001nk,Abbiendi:2003ji} had already made it nearly impossible to realize a low enough thermal relic density via their $t$-channel exchange.  The advent of slepton searches at the LHC has only strengthened this conclusion \cite{ATLASSlepton,CMSSlepton, Choudhury:2013jpa}.  But not only are the collider limits on the stau somewhat weaker, \cite{Heister:2001nk,Abdallah:2003xe,Abbiendi:2003ji}, the relatively large tau mass can lead to novel effects in annihilation, particularly at large $\tan \beta$.

In Sec.~\ref{stauexch} we outline preliminaries related to the stau-mediated annihilation process.  The following two sections discuss two distinct scenarios where the light stau can dominate the early universe annihilations.  In the first, the mostly bino dark matter has a non-trivial higgsino fraction (see Ref.~\cite{Calibbi:2013poa} for some recent discussion of this scenario).  In this case, the $\tan \beta$ enhanced Yukawa couplings of the stau can provide efficient enough annihilation.  The prospects for direct detection of dark matter in this case are relatively optimistic, as we will review.  In the second, large stau mixing is essential in producing a large enough ($s$-wave) annihilation of the neutralinos.  While direct detection may be unobservable in the near-term future in this case, it is likely that the staus could have an impact on the Higgs to diphoton branching ratio.  Both cases are imminently testable.

\section{Stau Exchange}\label{stauexch}
The neutralino is defined as 
\be
\chi = \tilde{B} N_{11} +\tilde{W} N_{12}+\tilde{H}_dN_{13}+\tilde{H}_uN_{14}\;,
\ee
where $N_{ij}$ are the components of the diagonalization matrix for the neutralino, (see  Eq.~(\ref{eqn:neutralinoexpansion}) below).
As a first step towards understanding the contribution of stau-exchange to early universe annihilation, we examine the couplings of the stau to the neutralino~\cite{Rosiek:1995kg}:
\begin{eqnarray}
g_{\chi_1^0 \tilde{\tau}_1\tau_L}&\equiv&  g_{L_1} = \frac{\sqrt{2}}{v} \left(M_Z\cos\tau (N_{12} c_W +s_W N_{11})- m_{\tau }\frac{\sin\tau}{\cos\beta} N_{13} \right),  \label{eq:couplingsL} \\
g_{\chi_1^0 \tilde{\tau}_1\tau_R}&\equiv& g_{R_1}=-\frac{\sqrt{2} }{v}\left(2 M_Z \sin\tau  s_W N_{11}+m_\tau \frac{\cos\tau}{\cos\beta} N_{13} \right), \label{eq:couplingsR}
\end{eqnarray}
where $v=246$ GeV is the Higgs vacuum expectation value (vev), $s_W$ ($c_{W}$) is the sine (cosine) of the weak mixing angle, and $\cos\tau$ is the stau mixing angle:
\begin{equation}
\cos\tau=
\frac{-m_\tau(A_\tau-\mu \tan\beta)}{\left[m_\tau^2 (A_\tau-\mu \tan\beta)^2+\left(m_{L_3}^2+m_\tau^2-m_{\tilde{\tau}_1}^2-\frac{1}{2} M_Z^2 c_{2W} \cos 2 \beta \right)^2\right]^{1/2}}.
\label{eqn:ctau}
\end{equation}
Here, $A_{\tau}$ is the tau trilinear coupling which we will always assume to vanish for simplicity; $\mu$ is the higgsino mass parameter; $m_{L_3}$ is the left-handed third generation slepton soft mass and  $m_{\tilde{\tau}_1}$ is the lighter stau mass eigenvalue. For completeness and to clarify conventions, we include the stau mass matrix in Appendix~\ref{Astau}. The couplings of the heavy stau to the neutralino are given by replacing $\cos\tau \to - \sin\tau$ and $\sin\tau\to \cos\tau$ in the above. In each coupling, the first term is due to interactions with gauginos, whilst the final terms in Eqs.~(\ref{eq:couplingsL}),(\ref{eq:couplingsR}) are due to the coupling to the higgsino.  These last terms can be enhanced at large $\tan \beta$. 

The  thermal annihilation cross section can be expanded as
\begin{equation}\label{eqn:ab}
\langle \sigma v \rangle_{x} = a+6\frac{b}{x}\;,
\end{equation}
with  $x=m/ T$. $a$ is the $s$-wave piece; $b$ is the $p$-wave contribution.  We used {\tt CalcHEP\_3.4}~\cite{Belyaev:2012qa, Pukhov:1999gg, Belanger:2004yn} to compute the neutralino annihilation cross-section to a pair of taus via the  $t$-channel exchange of the staus. Ignoring sub-leading  contributions to the $p$-wave cross-section (see below), the coefficients $a$ and $b$ of Eq.~(\ref{eqn:ab})  are given by  
\begin{eqnarray}\label{eqn:stau_exchange}
a &=&\frac{m_{\chi }^2}{8 \pi } \left(\frac{g_{L_1}  g_{R_1} }{ \left(m_{\tilde{\tau }_1}^2+m_{\chi }^2\right)} +\frac{g_{L_2}  g_{R_2} }{ \left(m_{\tilde{\tau }_2}^2+m_{\chi }^2\right)}\right)^2\;,
\nonumber \\
b&\approx& \frac{m_{\chi }^2}{48 \pi }\left[\frac{ (g_{L_1}^4 +g_{R_1}^4) \left(m_{\tilde{\tau }_1}^4+m_{\chi }^4\right) }{   \left(m_{\tilde{\tau }_1}^2+m_{\chi }^2\right)^4}+\frac{ (g_{L_2}^4 +g_{R_2}^4) \left(m_{\tilde{\tau }_2}^4+m_{\chi }^4\right) }{   \left(m_{\tilde{\tau }_2}^2+m_{\chi }^2\right)^4}\right] \;.
\end{eqnarray}

The possibility of having both $g_{R_i}$ and $g_{L_i}$ non-vanishing changes the picture qualitatively with respect to the pure bino and unmixed stau case ($N_{i1} = \delta_{i1}, \cos{\tau}=0$).  With both $g_{R_i}, \,g_{L_i} \neq 0$ an $s$-wave contribution to annihilation appears.  So, even though  $g_{R_i}$ dominates over much of the parameter space, the value of $g_{L_i}$ plays a crucial role. This is in contrast to smuon and slectron exchange, where both higgsino couplings and mixing are expected to be small, and the $s$-wave contribution is negligible. Additionally, the $s$-wave contribution to the annihilation cross-section is only proportional to $g_{L_i}g_{R_i}$. Therefore, the $p$-wave contribution is  relevant  only when one of the couplings is clearly suppressed. This justifies our neglecting all but the $g_{L_i}^4$ and  $g_{R_i}^4$ terms in the $p$-wave piece.

As mentioned in the introduction, we are interested in the case where the LSP is predominantly a bino (perhaps with a small higgsino admixture). In the pure bino limit, its couplings reduce to: $g_{L_1}\to \sqrt{2} \cos\tau s_w m_z/v$ and $g_{R_1}\to -2\sqrt{2} \sin\tau s_w m_z/v$. We see that for the most part, the lightest mostly right-handed stau exchange will dominate the annihilation cross-section.  However,  stau mixing effects can be important. For sufficient stau mixing, the dominant contribution would come from the $s$-wave piece in the annihilation cross-section (as $\cos\tau\sin\tau$ would be large). 

To better understand deviations from the pure bino limit, an expansion in $M_Z/\mu$ is instructive. Following Ref.~\cite{Arnowitt:1995vg}, in the limit $\mu$, $M_2 >> M_1, M_Z$ 
\begin{eqnarray}
N_{11} & \approx &1, \label{eqn:neutralinoexpansion}\\\
N_{12} & \approx &
-{1\over2}{M_{Z}\over\mu}{\sin2\theta_W\over{(1-M_{1}^{2}/\mu^{2})}}{M_{Z}
\over{M_{2}}-M_{1}}\left[\sin2\beta +
{M_{1}\over\mu}\right], \nonumber \\
N_{13} & \approx &{M_{Z}\over\mu}{1\over{1-M_{1}^{2}/\mu^{2}}}
\sin\theta_{W}\sin\beta\left[1+{M_{1}\over\mu} \cot\beta\right] ,
\nonumber \\
N_{14} & \approx &-{M_{Z}\over\mu}{1\over{1-M_{1}^{2}/\mu^{2}}}
\sin\theta_{W} \cos\beta\left[1+{M_{1}\over\mu} \tan\beta\right] 
\nonumber.
\end{eqnarray} 
Here, $M_{1}$ and $M_{2}$ represent the bino and wino soft masses respectively. From these expressions, there are three main take home messages.  First, the higgsino fractions decouple as $M_{Z}/\mu$.  Second, at large $\tan \beta$ and not too small $\mu$ we see that the down-type higgsino fraction exceeds that of the up-type higgsino fraction by a factor of $\mu/M_{1}$. 
Here, this ratio will always be large, as we are interested in light neutralinos (lighter than a light stau), with masses $m_{\chi} \approx M_{1} \lsim  M_{Z}$, and $\mu$ is constrained by direct collider searches  to be larger than $\mu \gsim 250$ GeV.  Finally, the wino fraction is also typically small.  In our numerical studies we take  $M_2 =400$ GeV $>> M_1$, in which case it plays a minor role.
For completeness we present the full neutralino mass matrix and  mixing angles in Appendix~\ref{Aneut}.

Using the approximation presented in Eq.~(\ref{eqn:neutralinoexpansion}), we see that the higgsino coupling to the stau would be proportional to $\tan\beta/\mu$. This then implies that even if the lightest stau is completely right-handed,  adding a small higgsino fraction to the almost bino neutralino could produce a significant $g_{L_i}g_{R_i}$ as long as $\tan\beta$ is sufficiently large. If this is the case,  the annihilation cross-section would again be dominated by its $s$-wave contribution.

In terms of the effective thermal annihilation cross section in Eq.~(\ref{eqn:ab}),(\ref{eqn:stau_exchange}), the relic density is given by:
\begin{equation}
\Omega h^2 = \frac{8.7\times 10^{-11} \text{GeV}^2}{\sqrt{g_*} \int^\infty_{x_F} \langle \sigma v\rangle_{x} x^{-2}}  =\frac{  8.7\times 10^{-11} \text{GeV}^2  x_F}{\sqrt{g_*}(a+3b/x_F) },
\end{equation}
where $g_*\sim 86.25$ are the degrees of freedom and $x_F\sim25$ at freeze out. We will generally require that the relic density be within approximately 3$\sigma$ of the experimental observation given in Ref.~\cite{Ade:2013zuv},
\be
 \Omega h^2 = 0.12 \pm 0.01.
\ee
Our choice of a relatively large error bar is also consistent with our choice of dropping some terms in the relic density calculation, which should give correction of $\lesssim 10\%$ in the relevant regions.   In the next two sections we will discuss two distinct scenarios where the neutralino relic density arising in the ``bulk" stau region may be consistent with the observed one.

But before doing this, we comment on the indirect detection situation.  For light neutralino masses with substantial $s$-wave annihilation, existing experimental bounds  can be relevant. A published bound from the FERMI collaboration coming from an examination of dwarf spheroidals \cite{Ackermann:2011wa} indicates that dark matter annihilating to tau leptons with thermal cross section is excluded up to masses of roughly 40 GeV.  An independent analysis \cite{Mazziotta:2012ux} gives similar results, $m_{\chi} > 30$ GeV.  Because our dark matter indeed annihilates dominantly via  $s$-wave (typically 90\% or greater of the annihilation is from $s$-wave in this low mass region), these bounds are relevant, so we keep in mind throughout that light dark matter with masses less than $\approx 25$ GeV is disfavored.  Moreover, these bounds are expected to improve.  Indeed, while this is not the focus of this work, projections from \cite{Morselli:2013wqa} indicate that the entire neutralino mass region up to 100 GeV should be probed by FERMI LAT on the 10 year time scale, assuming that astrophysical dark matter profiles are not too unfavorable. 

\section{Higgsino Doping}

If $\mu$ is not too large, there is a non-trivial higgsino component in the neutralino dark matter.  Further, the down-type higgsino has $\tan \beta$ enhanced couplings to the stau, so at large $\tan \beta$ this coupling can play an especially significant role. It is thus possible to realize the proper thermal relic abundance without any significant mixing in the stau sector $(\cos\tau=0$).  Taking the limit of vanishing mixing, and keeping the leading contribution in the neutralino mixing matrices, see Eq.~(\ref{eqn:neutralinoexpansion}), we find:
\begin{eqnarray}
g_{L_1} &\approx& \frac{\sqrt{2}}{v} M_{Z} s_{W}  \left( \frac{m_{\tau} \tan \beta}{\mu} \right),\\
g_{R_1} &\approx&-\frac{2 \sqrt{2}}{v} M_{Z} s_{W}.
\label{eq:BHcoups}
\end{eqnarray}
Note $|g_{L_1}/g_{R_1}| \approx m_{\tau} \tan \beta/(2 \mu)$, which is typically much less than one, and as we mentioned in the previous section, $g_{R_1}$ is the dominant coupling. However, since both couplings are non-zero, the $s$-wave contribution to $ \langle \sigma v \rangle$ exists and controls the relic density. Due to the $\tan\beta$ dependance of $g_{L_i}$,  one can realize the required relic density by dialing the values of $\tan\beta$ for a given neutralino mass, $m_\chi$, and $\mu$, which fix the  bino and down-type higgsino fractions of the neutralino.

\begin{figure}
\begin{center}
\includegraphics[width=0.6\textwidth]{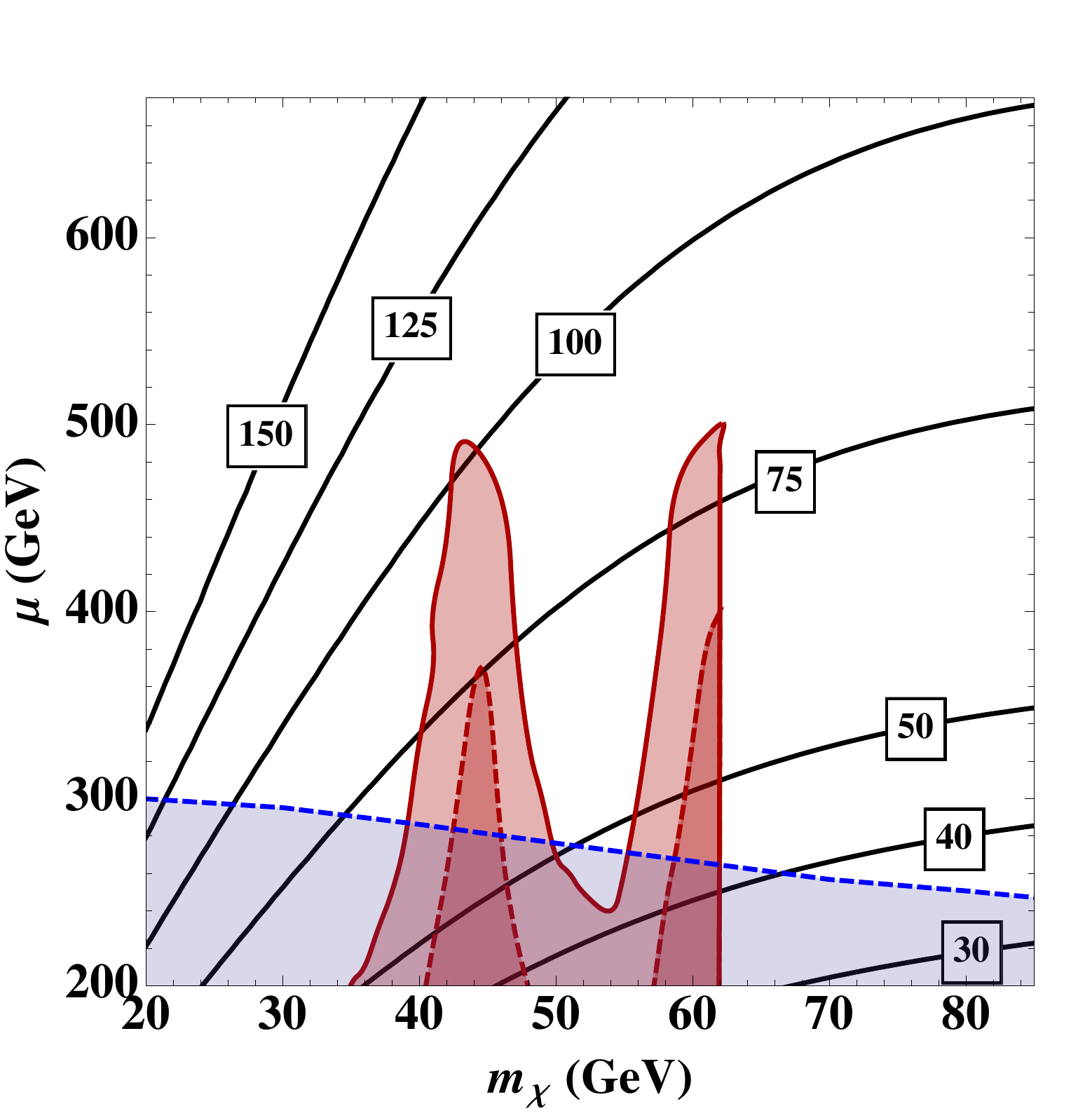}  
\end{center}
\caption{Contours of $\tan\beta$ giving rise to $\Omega h^2=0.12$ in the $m_\chi-\mu$ plane, for fixed $m_{\tilde{\tau}_1} = 95$ GeV. The contours are drawn using the approximation of solely stau-mediated annihilation. Also superimposed are red shaded regions denoting where final states other than $\tau\tau$ become relevant. The red solid/dashed boundaries correspond to when the $\tau\tau$ final state is less than 90\% and 50\% respectively of the total annihilation cross-section.  The region below the blue dashed line is excluded by an approximate recasting of CMS limits from Ref.~\cite{CMSEWK}, see the text for details.  The white region is the bulk stau region favored by the present analysis.}
\label{fig:BHDM}
\end{figure}

The values of $\tan\beta$ obtained by requiring $\Omega h^2=0.12$ are shown in Fig.~\ref{fig:BHDM} in the $\mu-m_\chi$ plane.   To construct these contours, we have utilized the analytic expression of Eq.~(\ref{eqn:stau_exchange}), fixing $m_{\tilde{\tau}_1} = 95$ GeV. $m_{\tilde{\tau}_2}$ is fixed to 10 TeV, effectively decoupling it from the analysis in this case. The effect of increasing the light stau mass on the relic density is to decrease $\langle \sigma v\rangle$. This can in principle be compensated for by increasing the values of $\tan\beta$ portrayed in Fig.~\ref{fig:BHDM}, but eventually uncomfortably large values of $\tan \beta$ are required. 

To ensure that limiting ourselves to light stau exchange is a reasonable approximation, we  also examined this region using {\tt{micrOMEGAs\_3.2}} \cite{Belanger:2013oya}. In the parameter space in Fig.~\ref{fig:BHDM} that is shaded red, the $Z$-pole and $h$-pole can contribute significantly. In the darkest region, these channels contribute at least 50\% to the early universe annihilation of neutralinos. In the lighter red region, they contribute at least 10\% to the annihilation.  Away from these red regions, the analytic stau exchange is an excellent approximation.  As the value of $\mu$ increases, the higgsino fraction of the lightest neutralino becomes smaller (again, see Eq.~(\ref{eqn:neutralinoexpansion})). To compensate, larger values of $\tan\beta$ are required. 

We also see the dependence of $ \langle \sigma v \rangle$ on the neutralino mass reflected in these contours. Examining the cross section, we find  (for fixed $m_{\tilde{\tau}_1}$) at small $m_{\chi}$, $\langle \sigma v \rangle$ grows approximately linearly with the neutralino mass.  However, as $m_\chi$ approaches $m_{\tilde{\tau}_1}$, the cross-section flattens out. This is the origin of the turn over in the $\tan \beta$ contours at the larger neutralino masses.

At large values of $\mu\tan\beta$ with light staus, there may exist a charge breaking vacuum deeper than the electroweak one, which  could lead to the instability of the electroweak vacuum at the timescale of the lifetime of the universe. To avoid such an instability, values of $\mu\tan\beta$ would have to be bounded from above~\cite{Hisano:2010re, Kitahara:2012pb, Carena:2012mw,Kitahara:2013lfa}. However, this bound is significantly alleviated when $\tilde{\tau}_2$ is much heavier than $\tilde{\tau}_1$. In fact for $m_{\tilde{\tau}_2} \gtrsim 2$ TeV, these considerations do not constrain our parameter space of $\mu$ and $\tan\beta$~\cite{Kitahara:2013lfa}

Also, collider constraints have begun to impinge upon this region. CMS has presented preliminary results in a trilepton search that directly considers the possibility that neutralinos and charginos cascade via an on-shell stau \cite{CMSEWK}.  These bounds depend on the neutralino mass, but roughly indicate limits on chargino masses of approximately 300-350 GeV.  However, we expect a somewhat reduced cross section with respect to the reference model utilized there, which assumes degenerate wino-like $\chi_2^0$ and $\chi^{\pm}_{1}$.  Instead, we have approximately degenerate higgsino-like $\chi_{2,3}^0$ and $\chi^{\pm}_{1}$. While we leave detailed recasting for future work, we do make an estimate of how the limits degrade for our model. 

As in the CMS reference model, production is dominated by the $s$-channel exchange of the $W$, for us to $\chi_{2,3}^0$ and $\chi^{\pm}_{1}$.  For a given mass, the production cross section for each neautralino in our model~($\chi_{2,3}^0 \approx$ higgsino)  is reduced by roughly a factor of four with respect to the reference model ($\chi_{2}^0 \approx$ wino).  Together these two states would have half the production cross section of the reference model for an identical mass.  Moreover, the branching ratio to the $\tau$ rich final state depends on the precise values of $m_{\chi}, \mu$ and $\tan \beta$.  We compute the relevant branching ratio using {\tt SUSY-HIT}~\cite{SUSYHit} for points in the $\mu, m_{\chi}$ plane at values of $\tan \beta$ that would reproduce the proper relic density.\footnote{We have checked that the subdominant decays $\chi_{2,3}^0 \rightarrow Z(h) \, \chi_{1}^0$ give weaker bounds.}   We then adapt limits from Ref.~\cite{CMSEWK} by finding the value of $\mu$ which has a cross section$\times$ BR comparable to the one excluded at the limit. This forces $\mu \gtrsim 250-300$ GeV, with a weakening of this bound towards larger values of the neutralino mass\footnote{Due to acceptance effects, we might also expect a subdominant dependence on the stau mass.  In Ref.~\cite{CMSEWK} this is taken to be the average between the chargino mass and the LSP mass.}.   ATLAS has presented comparable results \cite{ATLASTAU} derived from searches for hadronic taus and missing energy in the context of a similar, but not identical simplified model. ATLAS has also interpreted this result in terms of a MSSM model \cite{ATLASTAU} which suggests comparable bounds on $\mu$ ($\gtrsim 300$ GeV) to our derived CMS bounds~(see also Ref.\cite{Calibbi:2013poa}). However, the large $1\sigma$ band around the expected limit suggests that the results are extremely sensitive to small fluctuations which could degrade the observed limit considerably.   So at present, we concentrate on the CMS bounds which appear to be more robust.  We denote this bound as a blue dashed line in Fig.~\ref{fig:BHDM}. The present relevance of (both) limits to the parameter space at hand indicates that direct searches for electroweak production of these neutralinos at the 14 TeV LHC should be a powerful probe of this model~(see also Ref.~\cite{Choudhury:2013jpa}). More detailed examination is left for future work.

In addition, the coupling of the Higgs boson to the neutralino is governed by its higgsino fraction. Since obtaining  the correct relic density requires a non-negligible higgsino component, one might worry that this coupling could become large and experimental limits on invisible decays of the Higgs boson might be relevant \cite{Calibbi:2013poa}. However, the dominant coupling is to the up-type higgsino fraction, which is suppressed by an additional factor of $M_{1}/\mu$.   The result is that in the region of parameters under study (after imposing the direct constraints on $\mu$ above), we get less than 10\% invisible branching ratio --  well below the current experimental limit of $\sim 20$\% \cite{Falkowski:2013dza, Giardino:2013bma}. The largest branching ratios occur for the smallest values of $\mu$ and might be be probed with the full 14 TeV LHC data set \cite{PeskinHiggs}.

In determining the maximum allowed $\tan \beta$, perturbativity of the bottom Yukawa is quite relevant.  We should keep in mind that there could be important modifications to the $b$ Yukawa due to loop corrections~\footnote{The tau Yukawa may be corrected in a similar fashion. However, in the parameter region we are investigating, $\epsilon_\tau$ is negligible, and thus such corrections are irrelevant.} parameterized by $\epsilon_b$, $y_b = \sqrt{2}m_b/[v \cos\beta (1+ \epsilon_b \tan\beta)]$~\cite{Hall:1993gn, Pierce:1996zz,Carena:1998gk,Carena:2002es,Guasch:2001wv}. Since  $\epsilon_b$ is generally positive, it reduces the value of $y_b$ for a given value of $\tan\beta$. This has the effect of allowing values of $\tan\beta$ larger than what is generally considered perturbative. On the other hand, $\epsilon_b$ is controlled dominantly by squark-gluino loops which are completely unrelated to the analysis at hand, and one could imagine these particles also being very heavy, in which case $\epsilon_b\sim0$. Even then, relaxing the requirement that the Yukawa coupling be perturbative all the way up to the GUT scale, it is possible to have $\tan\beta \sim 100$ for cutoff scales of $\sim 10^7$ GeV.

\subsection{Direct Detection}
The non-trivial higgsino fraction of the neutralino ensures a not too-small contribution to direct detection from Higgs boson exchange.  Moreover, the large $\tan \beta$ required to give the correct relic density indicates the possibility of large contributions from the heavy Higgs boson.  

The spin-independent elastic cross-section for a neutralino scattering off a heavy nucleus due to the exchange of both the heavy and light  Higgs is given by  
\bea\label{eq:sigSI}
\sigma_{SI} = \frac{4m_r^2}{\pi} \left[Z f_p + (A-Z) f_n\right]^2
\eea
where $m_r = \frac{m_N m_{\chi}}{m_N+m_{\chi}}$, $m_N$ is the mass of the
nucleus, $m_{\chi}$ is the neutralino mass, and~\cite{Carena:2008ue}\footnote{There is a typographical error in Ref.~\cite{Carena:2008ue} in the $a_d$ component which we have fixed in Eq.~(\ref{ad:eq}). }
\bea
f_{p,n} &=& \left(\sum_{q=u,d,s} f_{T_q}^{(p,n)} \frac{a_q}{m_q} + \frac{2}{27}
f_{TG}^{(p,n)} \sum_{q=c,b,t} \frac{a_q}{m_q} \right) m_{(p,n)}, \\
a_u &=& - \frac{g_2 m_u}{4m_W s_{\beta}} (g_2 N_{12} - g_1 N_{11}) 
\left[N_{13} s_{\alpha} c_{\alpha} \left( \frac{1}{m_h^2} - \frac{1}{m_H^2}
\right) + N_{14} \left(\frac{c^2_{\alpha}}{m_h^2} +\frac{s^2_{\alpha}}{m_H^2}
\right)\right], \label{au:eq} \\
a_d &=&  \frac{g_2 \bar{m}_d}{4m_W c_{\beta}} (g_2 N_{12} -
g_1 N_{11}) 
\left\{ N_{13} \left[\frac{s^2_{\alpha}(1-\epsilon_d /t_\alpha)}{m_h^2} +\frac{c^2_{\alpha}(1+\epsilon_d t_\alpha)}{m_H^2}
\right] \right. \nonumber \\
&& \qquad  \qquad  \qquad \qquad \qquad \qquad \left.+ N_{14} s_{\alpha} c_{\alpha} \left[ \frac{(1-\epsilon_d /t_\alpha)}{m_h^2} - \frac{(1+\epsilon_d t_\alpha)}{m_H^2}
\right] \right\} \label{ad:eq}. 
\eea 
Here,  $\alpha$ is the mixing angle for the CP-even Higgs; $m_h$ is the light Higgs, and $m_H$ is the heavy Higgs boson mass. $ \bar{m}_d\equiv m_d/(1+\epsilon_d t_\beta)$,  where for completeness we have included the $\epsilon_d$ loop corrections to the couplings of the Higgs to down type fermions. In our numerics below, the $\epsilon_{d}$ are set to zero.  $m_{(p,n)}$ is either the proton or the neutron mass. For their respective form factors for $\{u, d, s\}$, we use the default parameters used by {\tt{micrOMEGAs\_3.2}}~\cite{Belanger:2013oya,Beringer:1900zz}:
\bea
f_{T_q}^p = \{0.0153, 0.0191, 0.0447\}; \qquad \qquad f_{T_q}^n =\{0.011, 0.0273, 0.0447\}\;.
\eea
Further, $f_{TG}^{(p,n)} =1-f_{T_u}^{(p,n)}-f_{T_d}^{(p,n)}-f_{T_s}^{(p,n)}$.

 \begin{figure}
\begin{center}
\begin{tabular}{c c}
(i) & (ii)\\
\includegraphics[trim = 0mm 0mm 0mm 0mm, clip,width=0.48\textwidth]{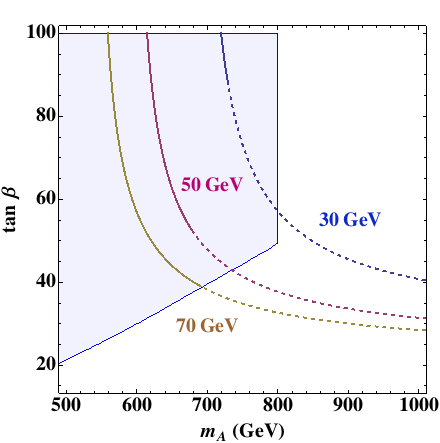}  &
\includegraphics[width=0.48\textwidth]{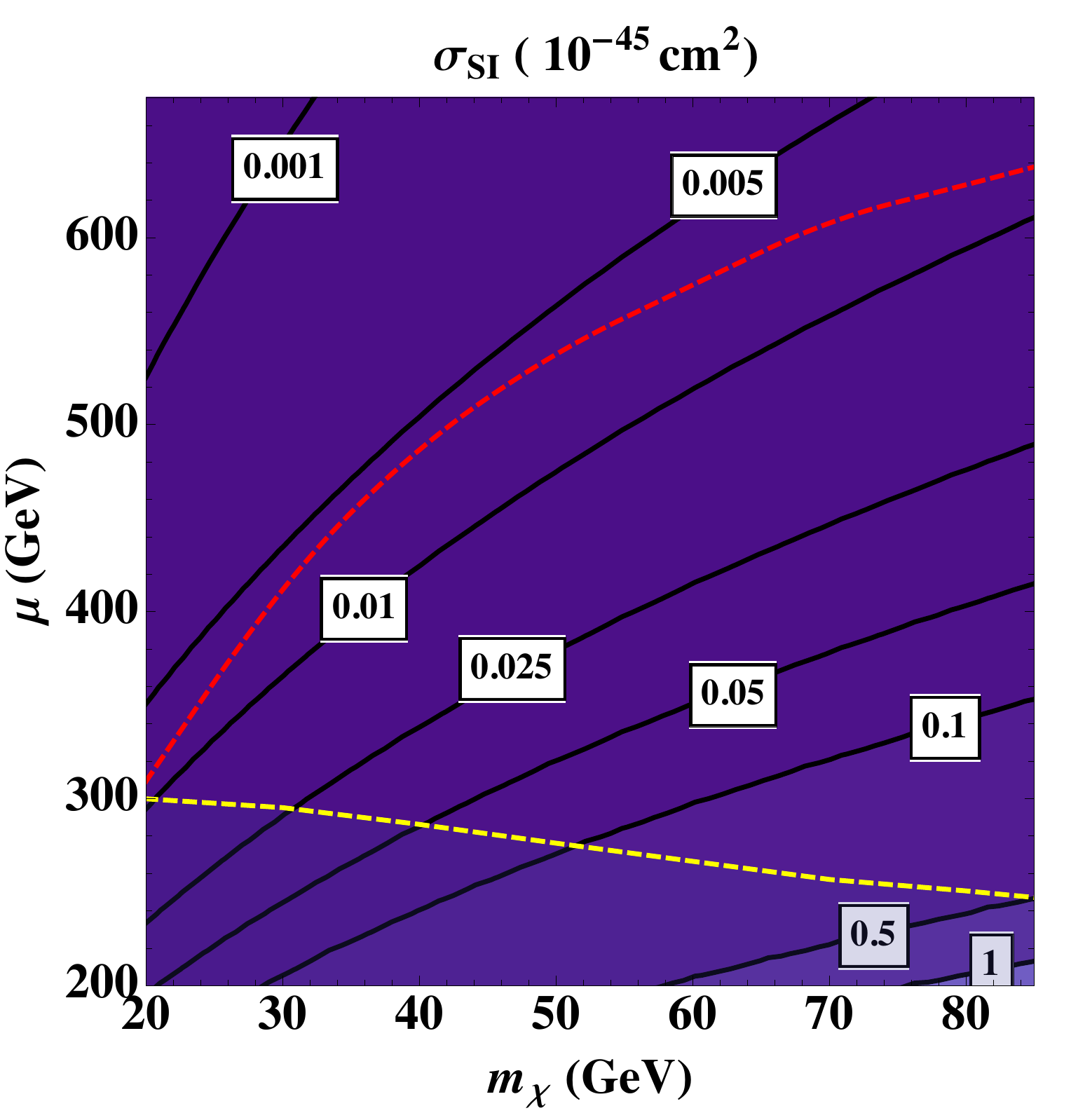} 
\end{tabular}
\end{center}
\caption{(i) Contours corresponding to three different neutralino masses, $m_\chi=30$, 50 and 70 GeV,  in the $\tan\beta-m_A$ plane for the current bounds from Xenon100. For any given point on the contour, the constraint from Xe-100 forces the value of $m_{A}$ to lie to the right of the contour.  The dashed portions of the lines denote values of $\mu$ which are likely excluded by direct searches for charginos \cite{CMSEWK}. The blue shaded region is excluded by direct searches for the heavy Higgs bosons of the MSSM via $H/A \rightarrow \tau \tau$ \cite{CMS-PAS-HIG-12-050}.  (ii) Contours showing $\sigma_{SI}$ with only the light Higgs contribution, i.e with $m_A\to \infty$. The dashed red line denotes regions below which Xenon1T would be sensitive. Below the dashed yellow line $\mu$ is disfavored by collider limits from Ref.~\cite{CMSEWK}. $m_{\tilde{\tau}_1}$ is fixed to 95 GeV in both panels.}
\label{fig:DD}
\end{figure}

In Fig.~\ref{fig:DD}(i) we explore the current interplay between direct detection constraints and collider constraints.  We again fix the lighter stau mass to 95 GeV. For each point in the $m_A$ vs $\tan \beta$ plane, a collider constraint can be applied directly, coming from the search for $A/H \rightarrow \tau \tau$. The current limits, coming from Ref.~\cite{CMS-PAS-HIG-12-050}  are shown as a shaded region. But how do these limits compare with considerations from the dark matter picture? The dark matter relic density depends on additional parameters, in particular $\mu$, $m_{\chi}$, and $m_{\tilde{\tau}_1}$. Due to the dominance of the $s$-wave contribution, the relic density is set through the product of $g_{L_{1}}$ and $g_{R_{1}}$ to an excellent approximation. Therefore, the relic density is merely a function of $\tan \beta$/$\mu$ (see Eq.~(\ref{eq:BHcoups})). So, for a fixed neutralino mass, $\mu$ varies approximately proportional to $\tan \beta$.  We use the values of $m_{\chi}$, $\tan \beta$ and the $\mu$ derived from the relic density constraint to calculate the direct detection cross section for various values of $m_{A}$. The contours for three different neutralino masses~($m_\chi=$ 30, 50 and 70 GeV) that saturate the current bound from Xe-100 \cite{Xenon100} are shown in Fig.~\ref{fig:DD}(i). For each neutralino mass, for a fixed value of $\tan \beta$ (and its derived $\mu$), values of $m_A$ to the right of the curve are allowed.  For each neutralino mass the solid portion of the line corresponds to favored values of $\mu$ -- the dashed portion is potentially excluded by direct searches for neutralinos.

A comment is in order on the shape of the curves in Fig.~\ref{fig:DD}(i). The spin independent direct detection contribution coming from the heavy  Higgs boson exchange has a piece enhanced by $\tan \beta$ and depends on the $\tilde{H}_d$ fraction of the LSP, which is inversely proportional to $\mu$. The result is that this contribution depends on the ratio $\tan \beta/ \mu$. But this is precisely the combination fixed by relic density considerations. Therefore, one might expect that fixing the relic density would precisely determine the direct detection cross section up to $m_A$, implying that the bounds would be independent of $\tan\beta$. In fact, this is why the curves in Fig.~\ref{fig:DD}(i) are approximately vertical at the largest values of $\tan \beta$. Deviations occur as $\tan \beta$ is decreased, and other contributions to direct detection become important. The leading  correction comes from the light Higgs boson exchange contribution to $a_d$ via the coupling to $\tilde{H}_{u}$, i.e. the second to last term of Eq.~(\ref{ad:eq}).

At present, the collider search is more constraining. And while the bounds from the direct search for the heavy Higgs are expected to improve~\cite{Arbey:2013jla}, future direct detection experiments will prove decisive. The crucial point is that the contribution from  the light (Standard Model-like) Higgs boson alone is also not negligible for the higgsino fractions at hand. For very large $m_A$ this would be the dominant contribution and  Eq.~(\ref{eq:sigSI}) reduces to:
\bea
\sigma_{SI}^h\approx \frac{g_2^2m_r^2  m_{(p,n)}^2}{4 \pi m_W^2 m_h^4 } \left(g_1 N_{11}-g_2 N_{12}\right)^2 \left(c_\beta N_{13}- s_\beta N_{14} \right)^2 (1-\frac{7}{9} f_{TG}^{(p,n)})^2. 
\eea
This contribution is shown as contours in Fig.~\ref{fig:DD}(ii). At each value of the neutralino mass and $\mu$, $\tan\beta$ is chosen such that $\Omega h^2=0.12$. Below the yellow dashed line $\mu$ would be disfavored by Ref.~\cite{CMSEWK}. The dashed red curve represents the reach of a 1 Ton-scale Xe experiment~\cite{Aprile:2009yh}. The 1 Ton-scale Xe experiment is expected to probe $\sigma_{SI}$ down to more than two orders of magnitude smaller than the present Xe-100 bounds.  All points below and to the right would be tested,
indicating that the majority of this region will be explored by such experiments,  even for $m_{A} \rightarrow \infty$.
 
The region which would not be excluded (at the top of the figure) corresponds to very large $\tan \beta$ (see Fig.~\ref{fig:BHDM}).  Because of the increased sensitivity of the 1 ton experiments, coupled with the large $\tan \beta$, it is possible that contributions from other heavy super partners could be relevant.  For example, the projected limit for $m_\chi=70$ GeV is $\sigma_{SI} \approx 7 \times  10^{-48} {\mbox{ cm}}^2$.  For $\mu=600$ GeV and $\tan\beta=90$ (consistent with relic density), even  2 TeV squarks give a non-negligible contribution, $\sigma_{SI}^{\tilde{q}} \approx 2\times 10^{-48} {\mbox{ cm}}^2$, comparable to the light Higgs boson contribution, $\sigma_{SI}^{h} \approx 7\times 10^{-48} {\mbox{ cm}}^2$.   Similarly, for these parameters a 3 TeV heavy Higgs boson will give a contribution of $\sigma_{SI}^{H} \approx 2\times  10^{-48} {\mbox{ cm}}^2$~\cite{Belanger:2013oya}. So, while the the contribution from the SM Higgs boson alone is not enough to ensure a signal at direct detection experiments,  even super partners with multiple TeV masses could contribute to observable rates.

\section{Gaugino dark matter and Stau Mixing}
A complementary possibility occurs when there is substantial mixing in the stau sector.  
From Eqs.~(\ref{eq:couplingsL}), (\ref{eq:couplingsR}), we see that if the mixing angle is near maximal, it is possible to have substantial $g_{L}$ and $g_{R}$, even with vanishing LSP  higgsino content. The couplings are then well approximated by
\bea
g_{L_1}&\approx& \frac{\sqrt{2}}{v} M_Z s_W \cos\tau,\nonumber\\
g_{R_1}&\approx&- \frac{2\sqrt{2}}{v} M_Z s_W \sin\tau. \label{eq:coup_staumix}
\eea
 
\begin{figure}
\begin{center}
\begin{tabular}{c c}
(i) & (ii)\\
\includegraphics[width=0.47\textwidth]{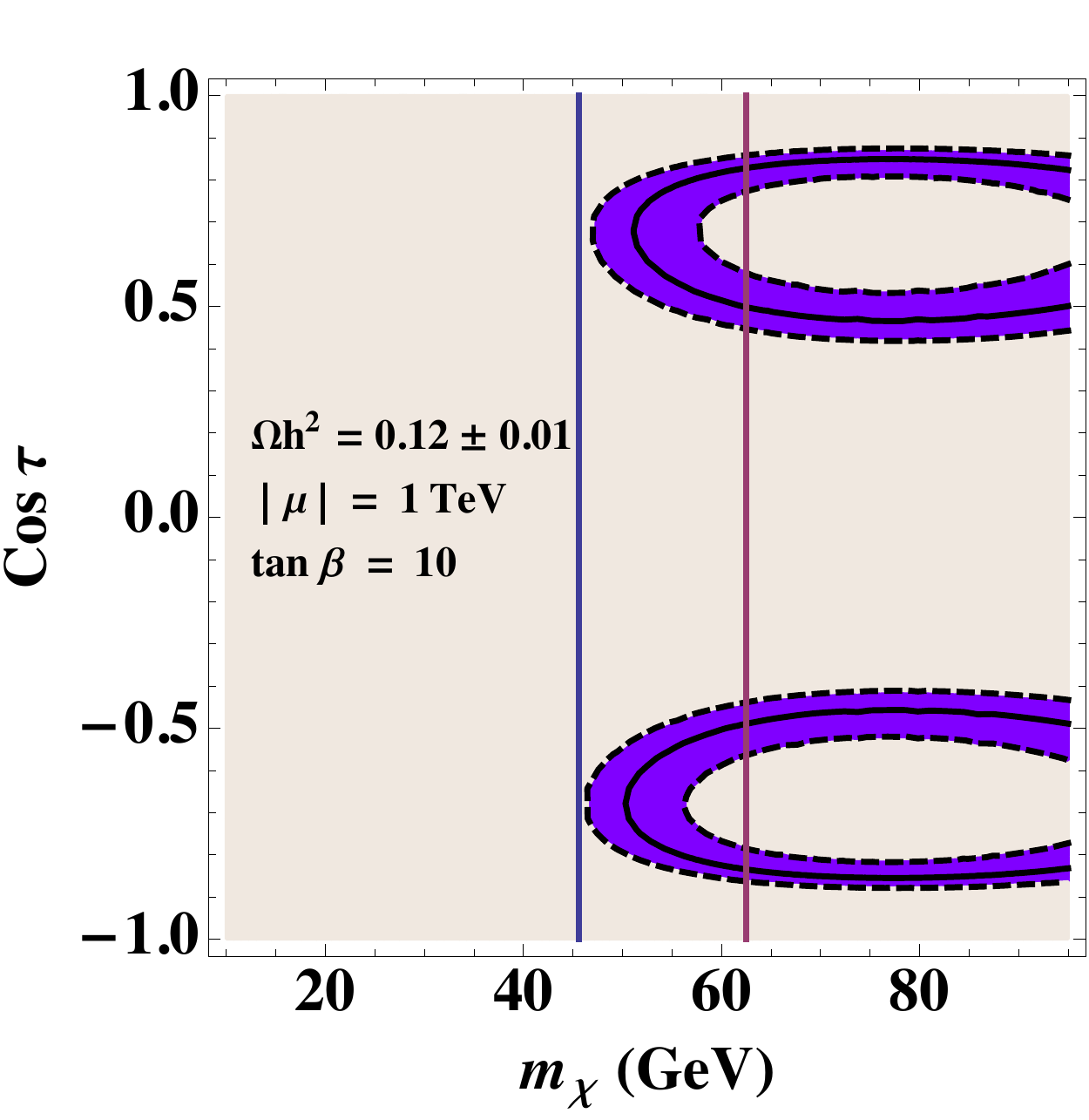} &
$\;\;\;$ \includegraphics[width=0.47\textwidth]{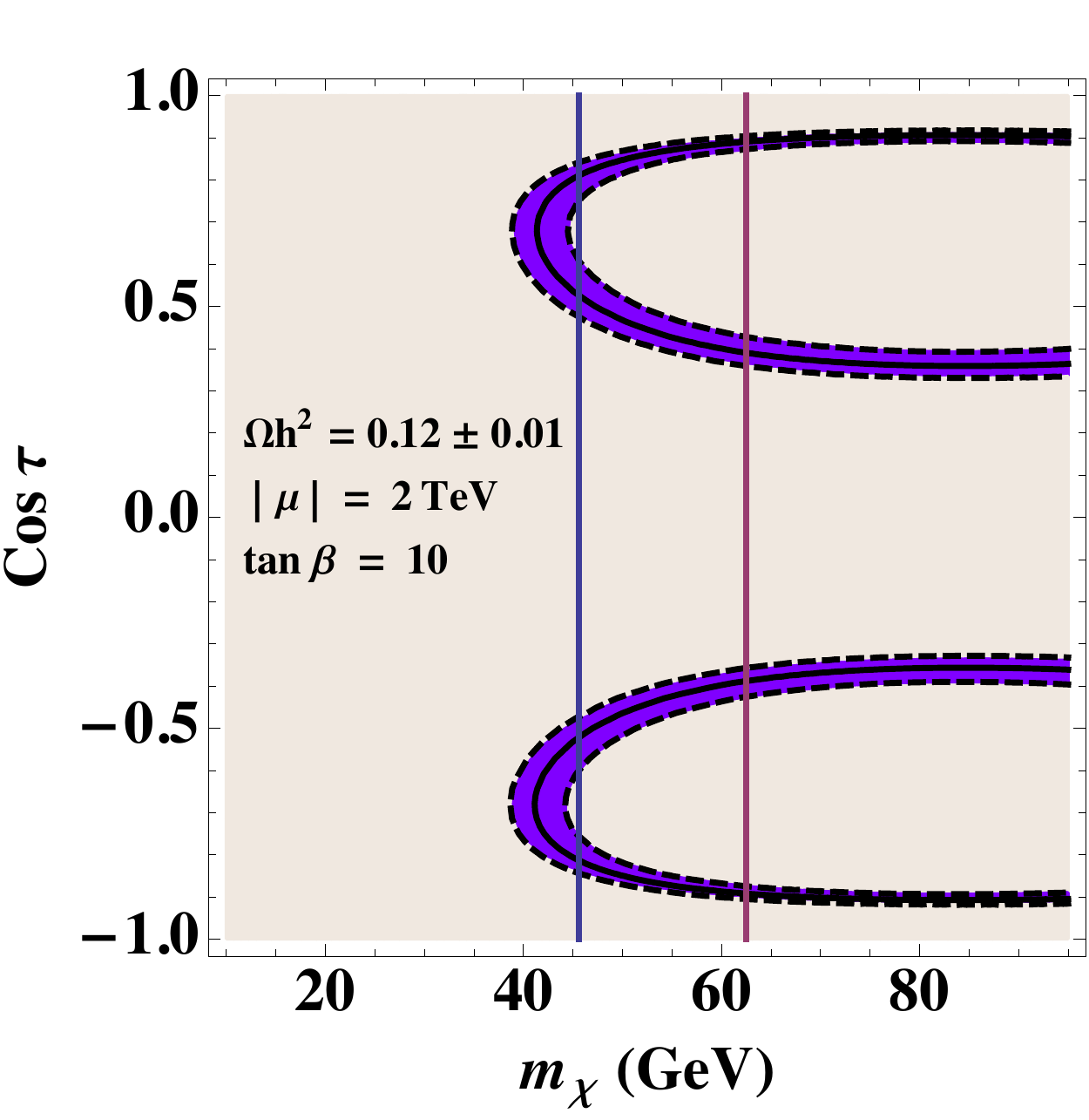}
\end{tabular}
\end{center}
\caption{Bands showing the  range $\Omega h^2 =0.12 \pm 0.01 $ for the relic density in the neutralino mass, stau mixing angle plane. Only the stau exchange is included as denoted in Eq.~(\ref{eqn:stau_exchange}).  Also shown are vertical lines at $m_{\chi}=$ $M_Z/2$ and $m_{h}/2$.  Right on top of these resonances, other channels may be important. (i): $|\mu| = 1$ TeV and $\tan \beta =10$. (ii): $|\mu| = 2$ TeV and $\tan \beta =10$.}
\label{fig:StauMixingOmh2}
\end{figure}

To explore this possibility, we fix $\mu$ to large values, so that the higgsino fraction will indeed be negligible. We then vary the neutralino mass and the stau mixing angle and plot contours of constant relic density, see Fig.~\ref{fig:StauMixingOmh2}.  We include only the exchange of the staus as given in Eq.~(\ref{eqn:stau_exchange}). The lighter stau mass is fixed to 95 GeV, and since we fix the value of $\mu$ and $\tan\beta$, the mass of the heavier stau is varied as the mixing angle is varied. The mass of ${\tilde{\tau}_2}$ for a given value of the mixing angle is 
\be
m_{\tilde{\tau}_2}^2 = m_{\tilde{\tau}_1}^2 + \left|\frac{m_\tau(A_\tau-\mu\tan\beta)}{\sin 2 \tau}\right|.
\ee

In Fig.~\ref{fig:StauMixingOmh2} we show plots for two choices of $\mu$ and $\tan \beta$: (i) $|\mu|=1 $ TeV, $\tan\beta=10$ and (ii) $|\mu|=2 $ TeV, $\tan\beta=10$. With our sign convention for the stau mixing angle, the sign of $\cos\tau$ is determined by the sign of $\mu$. We see that satisfying the relic density constraint via mixing in the stau sector requires almost maximal mixing. Otherwise, the annihilation cross-section is too small.  In addition, even at maximal mixing, for $m_\chi\lesssim 35$ GeV, it is not possible to realize a consistent relic density in this scenario. Also, if one increases the mass of the lightest stau, the effect on the relic density is significant. In this scenario, unlike the case of the bino-higgsino mixed neutralino, there is no parameter analogous to $\tan \beta$ that can be adjusted arbitrarily to increase the annihilation cross-section. Eventually, even for maximal stau mixing, the correct relic density cannot be obtained.  We checked numerically that for $m_{\tilde{\tau}_1}\gtrsim115$ GeV, it is impossible to reproduce the relic density for any neutralino mass in this case.

Note that for the smaller choice of $\mu$, i.e. Fig.~\ref{fig:StauMixingOmh2} (i),  the range of allowed neutralino mass is further limited. This is due to a larger destructive contribution from the heavier stau in this case, preventing efficient enough annihilation.   As mentioned before, the coupling of the heavy stau to the neutralino is given by the expressions in Eq.~(\ref{eq:couplingsL}),(\ref{eq:couplingsR}), but with $\cos\tau\to -\sin\tau$ and $\sin\tau\to \cos\tau$.  This sign ensures destructive interference between the contributions from the heavy and the light stau to the dominant $s$-wave contribution which is proportional to $g_L g_R$. The cancelation is more pronounced for the smaller values of $\mu\tan\beta$ since that corresponds to the lighter masses for $\tilde{\tau}_2$ for a given mixing angle. From Fig.~\ref{fig:StauMixingOmh2}(i) we see that having a light $m_{\tilde{\tau}_2}$ restricts the range of neutralino masses where the relic density constraint may be fulfilled to be larger than $\sim$ 45 GeV. 

Right at the $Z$ and the light Higgs boson resonances, one may still obtain the correct relic density even at very large values of $\mu$. However, because the higgsino fraction is suppressed by the large $\mu$, the couplings to both the $Z$ and $h$ are quite small, so the neutralino mass must be tuned quite precisely to resonance. We denote these resonances by lines at $m_\chi=M_Z/2$ and $m_h/2$.

Thus, even for a vanishing higgsino component in the LSP (large $\mu$) the correct relic abundance may be reproduced for sufficiently large mixing in the stau sector.  This scenario has strongly suppressed direct detection cross section.  However, we expect some consequences of this scenario for the Standard Model like Higgs boson.

\subsection{Higgs to Diphotons}

If the staus are indeed heavily mixed and light, they will impact the Higgs to diphoton branching ratio. As before, we will use analytic expressions for the Higgs-stau coupling and branching ratios to gain insight.
  
\begin{figure}
\begin{center}
\includegraphics[width=0.48\textwidth]{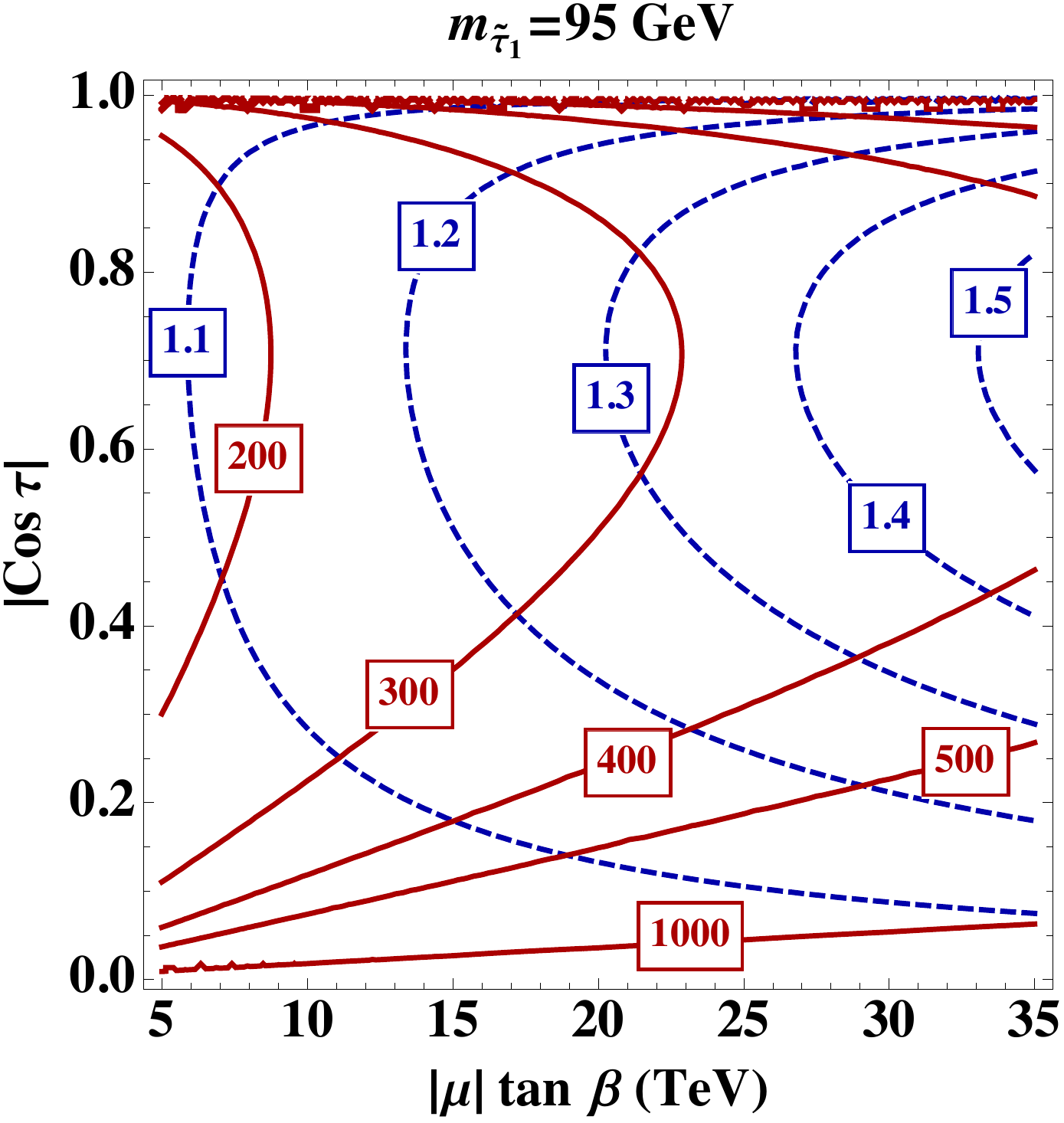}  
\end{center}
\caption{Ratio of the higgs boson branching ratio to diphotons compared to the SM value (denoted by blue dashed contours).  Also shown are values of the heavy stau mass, $m_{\tilde{\tau}_2}$ (GeV) (red solid lines). The mass of the lightest stau is fixed to 95 GeV.} 
\label{fig:Rgg}
\end{figure}

The new stau contributions can be simply included analytically via the relevant loop functions. The ratio of the Higgs to $\gamma\gamma$ width, including the effect of a light stau, to its SM value is given by~\cite{Carena:2012xa}\footnote{In our numerical work, i.e. Fig.~\ref{fig:Rgg}, we include the contribution of both $\tilde{\tau}$'s.}:
\begin{equation}\label{eq:rgg}
R_{\gamma\gamma}=\left|1+ \frac{v g_{h\tilde{\tau}_1\tilde{\tau}_1}A_0\left[\frac{4 m_{\tilde{\tau}_1}^2}{m_h^2}\right]}{m_{\tilde{\tau}_1}^2 \left\{2 \left[3 \left(\frac{2}{3}\right)^2 A_{1/2}\left[\frac{4m_t^2}{m_h^2}\right] +A_1\left[\frac{4 m_W^2}{m_h^2}\right]\right]\right\}}\right| ^2\;.
\end{equation}
The loop functions, $A_s[x]$, are defined in Appendix~\ref{Aloop} and the Higgs coupling to the lightest stau, in the decoupling limit, is given by
\bea
g_{h\tilde{\tau}_1\tilde{\tau}_1}&=&-\frac{2}{v} \left[m_z^2 \cos 2\beta ( \cos 2\tau s_w^2 -\frac{1}{2}\cos\tau^2 ) + 
   m_\tau^2 -\frac{1}{2}m_\tau (A_\tau - \mu \tan\beta)  \sin 2\tau \right], \label{eq:hstau}\\
   &\sim&  -\frac{1}{v} \left[\frac{1}{2}m_z^2-m_\tau (A_\tau - \mu \tan\beta) \sin 2\tau \right].\label{eq:hstauapp}
\eea

Numerically, the Standard Model contribution (i.e. the terms in the curly braces in Eq.~(\ref{eq:rgg})) to $R_{\gamma\gamma}$ is approximately (-13) and the loop factor $A_0\left[4 m_{\tilde{\tau}_1}^2/m_h^2\right] \approx 0.4$ for our benchmark point of $m_{\tilde{\tau}_1} = 95$ GeV.   Plugging these numerical values into  Eq.~(\ref{eq:rgg}),  light staus will significantly enhance  $R_{\gamma\gamma}$ if $|g_{h \tilde{\tau}_1\tilde{\tau}_1} | \gtrsim 80 $ GeV. This coupling is maximized at maximal mixing. So, with a light stau mass of 95 GeV, from Eq.~(\ref{eq:hstauapp}), $\mu\tan\beta \gtrsim 10$ TeV to provide more than 10\% enhancement.

From the previous section, we know that the relic density gives us two clues about the stau sector.  First, the mixing angle must be significant (see Fig.~\ref{fig:StauMixingOmh2}).  Second, if the heavy stau is too light then destructive interference precludes achieving the right relic density.
Indeed, numerical exploration indicates that the heavy stau should be at least approximately twice the light stau mass.  In fact, this is the reason for the narrower range of consistent neutralino masses in the left panel of Fig.~\ref{fig:StauMixingOmh2} compared with the panel with the larger value of $\mu \tan \beta$ -- for a given light stau mass, $\mu \tan\beta$, and the mixing angle,  $m_{\tilde{\tau}_2}$  is determined (with larger $m_{\tilde{\tau}_2}$ corresponding to larger $|\mu|\tan\beta$). To avoid a too large destructive interference, we find a large mixing indicates  $|\mu|\tan\beta \gtrsim 7$ TeV. This in turn implies a minimum $R_{\gamma\gamma}$ given a mixing angle.   

In Fig.~\ref{fig:Rgg} we plot contours of $R_{\gamma\gamma}$ (blue dashed lines) and $m_{\tilde{\tau}_2}$ (red solid lines) in the $|\mu|\tan\beta$-$|\cos\tau|$ plane. The lightest stau mass is again fixed to be 95 GeV.  Examining the plot, we see that the requirement of a thermal history (large mixing plus a minimum $\mu \tan \beta$, or alternatively a minimum $m_{\tilde{\tau}_2}$) indicates an enhancement of $R_{\gamma \gamma}$ of at least $\sim 10\%$.

But what is the largest $R_{\gamma \gamma}$ might be in this scenario?  To answer this question we must revisit the consequences of the vacuum stability constraint. This constrains  $\mu$ as a function of $\tan \beta$ and $m_{\tilde{\tau}_2}$. For a fixed $\tan\beta$, as the mass of $\tilde{\tau}_2$ is increased, the bound on $\mu\tan\beta$ is alleviated, which might lead one to believe that arbitrarily large enhancements in $R_{\gamma \gamma}$ were possible.   However, once we fix the lighter stau mass at 95 GeV, raising $m_{\tilde{\tau}_2}$ decreases the mixing angle. So,  even with the loosened vacuum stability bounds,  the net effect of raising the heavier stau mass is to \emph{decrease} the possible enhancement in $R_{\gamma\gamma}$, as shown in Ref.~\cite{Kitahara:2013lfa}.  Hence, the maximum possible value for $R_{\gamma\gamma}$ occurs at maximal mixing,  corresponding to approximately degenerate soft masses for the staus.  For this case, the stability of the vacuum is strongly dependent on $\tan\beta$~\cite{Carena:2012mw}. As an example, when $\tan\beta \sim 60$, vacuum stability demands  $|\mu|\tan\beta \lesssim 30$ TeV. On the other hand, for $\tan\beta\sim10$, one is constrained to have values of $|\mu| \tan\beta \lesssim 24$ TeV.  So, for any value of $\tan \beta$, there is a maximum possible value of $|\mu|\tan\beta$ and hence $R_{\gamma\gamma}$.  From Fig.~\ref{fig:StauMixingOmh2} we see that one needs to be at least in the region $0.4 \lesssim |\cos\tau| \lesssim 0.9$  to obtain an experimentally compatible relic density. Then, at moderate values of $\tan \beta$, choices consistent with vacuum stability yield enhancements of 30-40\%.

Interestingly, because large mixings are typical once the relic density is fixed, a definite value for $R_{\gamma\gamma}$ (along with perhaps a measurment of a light stau mass) points to a particular heavy stau mass.  For example, if one has $R_{\gamma\gamma}\sim 1.2-1.3$, then one would need to have the ${\tilde{\tau}_2}$ mass approximately in the 250-350 GeV window.    

We  also comment briefly on the impact of different choices of  $m_{\tilde{\tau}_1}$.
$R_{\gamma\gamma}$ has an inverse dependance on the lightest stau mass, but the value of  $m_{\tilde{\tau}_2}$ is mostly determined by $\mu\tan\beta/\sin2\tau$. Therefore, if $m_{\tilde{\tau}_1}$ is only raised slightly, so that  $m_{\tilde{\tau}_1} << m_{\tilde{\tau}_2} $ still obtains, then the $R_{\gamma\gamma}$  contours will quickly shift to the right (minimizing the relevant effect), whereas there will be a much less significant effect on the $m_{\tilde{\tau}_2}$ contours. As an example, if  $m_{\tilde{\tau}_1}\gtrsim 160$ GeV it would not be possible to obtain more than a few \%  enhancement in $R_{\gamma\gamma}$.

\section{Conclusions}\label{conc}
A light stau with mass $\sim 100$ GeV can help a predominantly bino LSP achieve the right thermal relic density.  Two distinct scenarios were explored, both relying on large $\tan \beta$.  In the first, a small down-type higgsino fraction in the LSP has a significant coupling to the stau.  The presence of this coupling allows significant $s$-wave annihilation.  The second also relies on  large $\tan \beta$, but can be realized in the limit of a pure bino.  Here, the key was  large mixing in the stau sector.  In this case, the lightest stau must have mass $\lesssim 115$ GeV.

Both scenarios should be tested soon, and allow ample opportunities for discovery. In the case with a non-trivial higgsino fraction, ton-scale detection experiments should easily probe the entire region.  Moreover, direct searches for the higgsinos at the LHC should prove promising.  In the pure bino case, direct detection is small, but the presence of two light staus with significant mixing can impact the properties of the Higgs boson; enhancements in the diphoton branching ratio with respect to the Standard Model value of up to 40\% are possible.

We have focused on two extremes -- a non-trivial higgsino fraction and large stau mixing.  It is also possible to imagine combinations of the two.  For example, one could perturb the mixed bino--higgsino case by adding stau mixing.  In this case, there typically is a partial cancellation in  some of the relevant couplings.  This can be compensated by going to larger values of $\tan \beta$. Experimentally, this case would realize a combination of the signals described.  In addition to possibly detectable higgsinos and direct detection signatures, it is possible to realize  (modest) deviations in the Higgs branching ratio to diphotons, see Fig.~\ref{fig:Rgg}.  

The possibility explored here is an exciting one where many light super partners are  kinematically accessible. Not only dark matter, but other particles crucial for the relic abundance story -- light  staus and charginos  are well within experimental reach.

\section*{Note added}
While this work was being completed, Ref.~\cite{Hagiwara:2013qya} appeared, which has substantial overlap in its discussion of $\tilde{\tau}$-mediated dark matter annihilation, and its implication for Higgs branching ratios.  Ref.~\cite{Hagiwara:2013qya} concentrates primarily on $m_{\chi} \approx 10$ GeV region and its consistency with indirect detection, but does not have a thermal history. 

\section*{Acknowledgements}
We thank  C.~Boehm, G.~Belanger, M.~Carena, J.~Kearney, and C.~Wagner for discussions.  KF, AP and NS are supported by DoE grant DE-SC0007859.  AP is also supported under CAREER grant NSF-PHY 0743315.  AP and NS thank the KITP, where this research was supported in part by the National Science Foundation under Grant No. NSF PHY11-25915.  NS thanks the Aspen Center for Physics and the NSF Grant \#1066293 for hospitality during the completion of this work. 
KF was supported in part by a Simons Foundation Fellowship in Theoretical Physics. KF, AP, and NS are supported by the Michigan Center for Theoretical Physics. 

\appendix
\section{Neutralino Mass Matrix}\label{Aneut}

The neutralino mass matrix is given by:

\begin{equation}
M_N=\left(
\begin{array}{cccc}
 M_1 & 0 & -\frac{v g_1c_{\beta}}{2} & \frac{v g_1 s_{\beta }}{2} \\
 0 & M_2 & \frac{v c_W g_1 c_{\beta}}{2 s_W } & -\frac{v c_W g_1 s_{\beta }}{2 s_W } \\
 -\frac{v g_1c_{\beta}}{2} & \frac{v c_W g_1 c_{\beta}}{2 s_W } & 0 & -\mu  \\
 \frac{v g_1 s_{\beta }}{2 } & -\frac{v c_W g_1 s_{\beta }}{2 s_W } & -\mu  & 0 \\
\end{array}
\right)
\end{equation}
where $g_1= 2 M_Z s_W/v$, $s_\beta \equiv \sin\beta$ and $c_\beta\equiv \cos\beta$.

The matrix diagonalizing $M_N$ is given by the matrix of its eigenvectors. The mass eigenstates are the roots to the following:
\begin{eqnarray}
\left(x-M_1\right) \left(x-M_2\right)  (x^2-\mu^2 )-M_Z^2 \left(x-M_1c_W^2-M_2 s_W^2\right) \left(2 \mu  s_{\beta }c_{\beta}+x\right) =0 \nonumber\\
&&
\end{eqnarray}

 In terms of the mass eigenstates, $m_{\chi_i}$, the eigenvectors are then exactly given by:

\begin{equation}
N_i=\frac{1}{\sqrt{C_i}}\left(
\begin{array}{c}
 \left(\mu ^2-m_{\chi _i}^2\right) \left(M_2-m_{\chi _i}\right)- M_Z^2 c_W^2\left(m_{\chi _i}+2 \mu  s_\beta c_\beta\right)\\
\\
- M_Z^2 s_W c_W \left(m_{\chi _i}+2 \mu  s_\beta c_\beta\right)\\
\\
 \left(M_2-m_{\chi _i}\right) \left(m_{\chi _i} c_\beta+\mu  s_\beta\right)M_Z s_W\\
\\
- \left(M_2-m_{\chi _i}\right) \left( m_{\chi _i}s_\beta+\mu c_\beta\right)M_Z s_W
\end{array}
\right)
\end{equation}
where
%
\begin{eqnarray}
C_i&&= M_Z^2c_W^2 \left(m_{\chi _i}+2 \mu  s_{\beta } c_\beta\right) \left[M_Z^2\left(m_{\chi _i}+2 \mu  s_{\beta } c_\beta\right)+2  \left(\mu ^2-m_{\chi _i}^2\right) \left(m_{\chi _i}-M_2\right)\right]\nonumber\\
&&\qquad \qquad+ \left(m_{\chi _i}-M_2\right)^2 \left\{M_Z^2 s_W^2 \left[ \left(m_{\chi _i}^2+\mu ^2\right)+4 \mu  m_{\chi _i} s_{\beta } c_\beta  \right]+ \left(m_{\chi _i}^2-\mu ^2\right)^2\right\}.
\end{eqnarray}

The diagonalizing matrix is then given by: $N=\{N_1,N_2,N_3,N_4\}$, where ${i=1,2,3,4}$ correspond to the bino, wino, higgsino-down and higgsino-up components respectively. If $\tan\beta$ is larger than about 10, and $m_\chi<<\mu$, the $\tan\beta$ dependance on the components is negligible. Further, for $M_2>>m_\chi$, the wino and the higgsino-up components are at most a few percent. 

\section{Stau Mass Matrix}\label{Astau}

The stau mass matrix is given by:

\begin{equation}
M^2_{\tilde{\tau}}=\left(
\begin{array}{cc}
m_{L_3}^2 + m_\tau^2 + D_L   & -\frac{h_\tau v}{\sqrt{2}} (A_\tau \cos\beta - \mu \sin\beta)   \\
-\frac{h_\tau v}{\sqrt{2}} (A_\tau \cos\beta - \mu \sin\beta)    &  m_{e_3}^2 + m_\tau^2 + D_R\\
\end{array}
\right),
\end{equation}
where $m_{L_3},m_{e_3}$ are the left and right-handed soft masses, $h_\tau$ is the tau Yukawa, $v=246$ GeV  is the Higgs vev and the D-terms are: $D_L= - M_Z^2 \cos 2 \beta (1-2 s_W^2)$ and $D_R=-M_Z^2 \cos 2\beta s_W^2$. 

We choose sign conventions for the mixing angle, $\cos\tau$, defined in the text in Eq.~(\ref{eqn:ctau}), such that we are consistent with those used by {\tt{SuSpect\_2.41}}.

\section{Loop Functions}\label{Aloop}
The loop functions we will need to calculate the Higgs boson to diphoton width are~\cite{Carena:2012xa}:
\begin{eqnarray}
A_0[x]&=&-x^2 \left[\frac{1}{x}-f\left[x^{-1}\right]\right] \\
\nonumber \\
A_1[x]&=&-x^2 \left[3 \left(\frac{2}{x}-1\right) f\left[x^{-1}\right]+\frac{2}{x^2}+\frac{3}{x}\right]\\
\nonumber \\
A_{1/2}[x]&=&2 x^2 \left[\left(\frac{1}{x}-1\right) f\left[x^{-1}\right]+\frac{1}{x}\right] \\
\nonumber\\
f[x]&=&\arcsin^2 \sqrt{x}
\end{eqnarray}

\begin{spacing}{1.1}
\bibliography{StauBib_AP}
\bibliographystyle{utphys}
\end{spacing}
\end{document}